\documentclass[pra,aps,twocolumn,epsfig,superscriptaddress,showpacs]{revtex4}
\usepackage{amsfonts}
\usepackage[dvips]{graphicx}

\usepackage{mathrsfs}
\usepackage[intlimits]{amsmath}

\begin{document}
\author{Jing-Ling Chen}
 \email{chenjl@nankai.edu.cn}
\affiliation{Theoretical Physics Division, Chern Institute of
Mathematics, Nankai University, Tianjin 300071, People's Republic of
China}
\author{Dong-Ling Deng}
 \affiliation{Theoretical Physics Division, Chern Institute of
Mathematics, Nankai University, Tianjin 300071, People's Republic of
China}
\author{Ming-Guang Hu}
\affiliation{Theoretical Physics Division, Chern Institute of
Mathematics, Nankai University, Tianjin 300071, People's Republic of
China}

\date{\today}



\title{Gisin's theorem for two $d$-dimensional systems based on the
Collins-Gisin-Linden-Masser-Popescu inequality}

\begin{abstract}
In this Rapid Communication, we show analytically that all pure
entangled states of two $d$-dimensional systems (qudits) violate the
Collins-Gisin-Linden-Masser-Popescu (CGLMP) inequality. Thus one has
the Gisin's theorem for two qudits.
\end{abstract}

 \maketitle

In $1964$, Bell published a celebrated inequality to show that
quantum theory is incompatible with local realism \cite{J.S.Bell}.
He showed that any kinds of local hidden variable theories based on
Einstein, Podolsky, and Rosen's notion of local realism
\cite{A.Einstein} should obey this inequality, while it can be
violated easily in quantum mechanics. Thus, Bell's inequalities made
it possible for the first time to distinguish experimentally between
local realism model and quantum mechanics. This applaudable progress
for the foundation of quantum mechanics has stirred a great furor,
and extensive earlier works on Bell inequalities have been done,
including the Clauser-Horne-Shimony-Holt (CHSH) inequality
\cite{J.Clauser} for bipartite system and the
Mermin-Ardehali-Belinskii-Klyshko (MABK) inequalities for
multipartite systems~\cite{MABK}. For more details about various
kinds of Bell inequalities one can refer to
\cite{R.F.Werner-T.paterek} and references therein. Now Bell
inequalities are widely used in many fields. Many experimenters use
the Bell inequalities to check whether they have succeeded in
producing entangled states~\cite{M. Nielsen}. Furthermore, Bell
inequalities are also used to realize many tasks in quantum
computation and quantum information, such as making the secure
quantum communication and building quantum protocols to decrease the
communication complexity \cite{C.Brukner-A.Acin}.

However, many problems are still open~\cite{N.Gisin}, such as: (a)
What are the most general Bell inequalities for $N$ qudits? (b)
Which quantum states violate these inequalities? and so on. For the
problem (b), Gisin presented a theorem in 1991 that any pure
entangled states of two spin-1/2 particles (qubits) violate the CHSH
inequality~\cite{Gisin-1991}. Soon after, Gisin and Peres provided a
more complete and simpler proof of this theorem for two arbitrary
spin-$j$ particles (i.e., the qudits) \cite{Gisin-Peres-1992}. They
have stressed an important topic indicating the relations between
quantum entanglement and Bell inequality. In their paper, they
constructed four observables, two for each subsystems and the
eigenvalues of these observables are $\pm1$. They proved that for
any entangled states, the correlations involved in the quantum
systems violate the CHSH inequality. The Gisin's theorem has also
been successfully generalized to three qubits. In 2004, Chen {\it et
al.} showed that all pure entangled states of a three-qubit system
violate a Bell inequality for probabilities~\cite{J.L. Chen-2004}.
This triumphant casus also reveals that the wisdom of Bell
inequality as a necessary and sufficient condition to quantify the
quantum entanglement is also held in a multi-particle system.
Despite all that, whether Gisin's theorem can be generalized for $N$
qudits or not remains open. There are two main difficulties: The
first is the problem (a) mentioned above, namely, before checking
the Gisin's theorem one has to firstly build a corresponding
$N$-qudit Bell inequality; The second is that Schmidt decomposition
is not valid for multipartite systems, consequently, the parameters
needed to describe a pure state of multi-particle systems grow
exponentially with the number of particles $N$ and the dimension
$d$. In fact, people don't know exactly how many Schmidt parameters
are needed to describe a pure state of multi-particle systems, even
for a three-qudit system.

There are renewed interests in studying the Gisin's theorem for a
two-qudit system by using various kinds of Bell inequalities. The
purpose of this Rapid Communication is to show analytically that all
pure entangled two-qudit states violate the CGLMP
inequality~\cite{2002Collins}. The brilliant idea of Gisin and Peres
was based on the CHSH inequality \cite{Gisin-Peres-1992}, and at
that time the tight Bell inequality for two qudits was not available
until the CGLMP inequality appeared in 2002. Our method is based on
the most recent CGLMP inequality, which is a natural generalization
of the CHSH inequality from two qubits to two qudits. From this
point of view, it is more natural to utilize the CGLMP inequality to
investigate the Gisin's theorem of two-qudit than the CHSH one. In
our method, we shall choose some special unitary transformation
matrices to show that all the entangled states violate the CGLMP
inequality. Since the CGLMP inequality is in the form of joint
probabilities, one only needs to perform some projective
measurements to calculate the joint probabilities, which may be more
convenient for experiments.

Let us  make a brief survey for the CGLMP inequality first. Consider
the standard Bell-type experiment: two spatially separated
observers, Alice and Bob,
share a copy of a pure two-qudit state
$|\psi\rangle\in\mathbb{C}^d\otimes\mathbb{C}^d$ on the composite
system. Suppose that Alice and Bob both have choices to perform two
different projective measurements, each of which can have $d$
possible outcomes. Namely, let $A_1$, $A_2$ denote the measurements
of Alice, $B_1$, $B_2$ denote the measurements of Bob, and each
measurement may have $d$ possible outcomes: $A_1, A_2, B_1, B_2=0,
\cdots, d-1$. Note that each observer can choose his/her
measurements independently of what the other distant observer does
(or has done or will do). Then any local variable theories must obey
the well-known CGLMP inequality~\cite{2002Collins}:
\begin{widetext}
\begin{eqnarray}\label{CGLMP}
I_d&=&\sum_{k=0}^{[d/2]-1}\left(1-\frac{2k}{d-1}\right)\;\{[P(A_1=B_1+k)+P(B_1=A_2+k+1)+P(A_2=B_2+k)+P(B_2=A_1+k)]\nonumber\\
&&-[P(A_1=B_1-k-1)+P(B_1=A_2-k)+P(A_2=B_2-k-1)+P(B_2=A_1-k-1)]\}\leq2.
\end{eqnarray}
\end{widetext}
Here $[x]$ denotes the integer part of $x$, and we denote the joint
probability $P(A_a=B_b+m)$ ($a, b=1, 2$) as
\begin{eqnarray}\label{jointp}
P(A_a=B_b+m)=\sum_{j=0}^{d-1}P(A_a=j, B_b=j-m),
\end{eqnarray}
in which the measurements $A_a$ and $B_b$ have outcomes that differ
by $m$ (modulo $d$). In the case of $d=2$, inequality (\ref{CGLMP})
reduces to the famous CHSH inequality. It was shown
in~\cite{Masanes} that the CGLMP inequality~(\ref{CGLMP}) is a facet
of the convex polytope generated by all local-realistic joint
probabilities of $d$ outcomes, that is, the inequality is tight.
This means that inequality~(\ref{CGLMP}) for two-qudit is optimal.
Our main result is the following Theorem.


\emph{Theorem.} Let $|\psi\rangle\in\mathbb{C}^d\otimes\mathbb{C}^d$
be a pure entangled two-qudit state, then it violates the CGLMP
inequality for any $d\ge 2$.

\emph{Proof.} The quantum prediction of the joint probability
$P(A_a=k, B_b=l)$ when $A_a$ and $B_b$ are measured in the state
$|\psi\rangle$ is given by
\begin{eqnarray}\label{jointpro}
P(A_a=k,B_b=l)&=&|\langle kl|U(A)\otimes U(B)|\psi\rangle|^2\nonumber\\
&=&\text{Tr}\{[U(A)^{\dagger}\otimes
U(B)^{\dagger}]\;\hat{\Pi}_k\otimes\hat{\Pi}_l\;\nonumber\\
&&\times [U(A)\otimes U(B)]|\psi\rangle\langle\psi|\},
\end{eqnarray}
where $U(A)$, $U(B)$ are the unitary transformation matrices, and
$\hat{\Pi}_k=|k\rangle\langle k|$, $\hat{\Pi}_l=|l\rangle\langle l|$
are the projectors for systems A and B, respectively.

We shall follow three steps to prove this theorem. First, the case
with $d=2$ is considered. The two-qubit state reads
$|\psi\rangle_{\rm
qubits}=\cos\theta_1|00\rangle+\sin\theta_1|11\rangle$. We choose
the unitary transformation matrices as
$$U(A)=\left(\begin{matrix}\cos\zeta_a&\sin\zeta_ae^{-i\phi_a}\\
\sin\zeta_ae^{i\phi_a}&-\cos\zeta_a\end{matrix}\right),$$
$$U(B)=\left(\begin{matrix}\cos\eta_b&\sin\eta_be^{-i\varphi_b}\\
\sin\eta_be^{i\varphi_b}&-\cos\eta_b\end{matrix}\right).$$
Substituting them into the inequality~(\ref{CGLMP}), and choosing
the following setting $\zeta_1=0$, $\zeta_2=\pi/4$, $\phi_1=0$,
$\phi_2=0$, $\varphi_1=0$, $\varphi_2=0$, we get
$I_2=\cos2\eta_1-\sin2\theta_1\sin2\eta_1+\cos2\eta_2+
\sin2\theta_1\sin2\eta_2\leq2\sqrt{1+\sin^2(2\theta_1)}$. The equal
sign occurs at $\eta_1=-\eta_2=-\tan^{-1}[\sin(2\theta_1)]$.
Obviously, the CGLMP inequality is violated for any $\theta_1\neq0$
or $\pi/2$. Second, we consider the case with $d=3$. The two-qutrit
state reads $|\psi\rangle_{\rm
qutrits}=\cos\theta_2(\cos\theta_1|00\rangle+\sin\theta_1|11\rangle)+\sin\theta_2|22\rangle$.
We choose the unitary transformation matrix of particle A as:
$U(A)=\cos\zeta_a|0\rangle\langle0|+\sin\zeta_ae^{-i\phi_a}|0\rangle\langle1|+\sin\zeta_ae^{i\phi_a}|1\rangle\langle0|
-\cos\zeta_a|1\rangle\langle1|+|2\rangle\langle2|$, or in the matrix
form:
\begin{eqnarray}\label{qutrit-unitary}
U(A)=\left(\begin{matrix}\cos\zeta_a&\sin\zeta_ae^{-i\phi_a}&0\\
\sin\zeta_ae^{i\phi_a}&-\cos\zeta_a&0\\
0&0&1
\end{matrix}\right).
\end{eqnarray}
The unitary transformation matrix $U(B)$ has the same form as
$U(A)$. Substituting them into the CGLMP inequality, and choosing
the following setting $\zeta_1=0$, $\zeta_2=\pi/4$, $\phi_1=0$,
$\phi_2=0$, $\varphi_1=0$, $\varphi_2=0$, we get
$I_3=\frac{1}{4}(2+3\cos2\eta_1-3\sin2\theta_1\sin2\eta_1+3\cos2\eta_2+
3\sin2\theta_1\sin\eta_2)\cos^2\theta_2+2\sin^2\theta_2\leq\frac{1}{2}
(1+3\sqrt{1+\sin^22\theta_1})\cos^2\theta_2+2\sin^2\theta_2$. The
equal sign occurs at $\eta_1=-\eta_2=-\tan^{-1}[\sin(2\theta_1)]$.
It is obvious that $1+3\sqrt{1+\sin^22\theta_1}$ is larger than $4$,
so the maximal value of $I_3$ is larger than $2$, which means the
CGLMP inequality is violated for any $\theta_1\neq0$ or $\pi/2$.
Finally, the case with $d\ge 4$ is considered. The state of two
qudits ($d\ge 4$) reads
\begin{widetext}
\begin{eqnarray}\label{states}
|\psi\rangle_{\rm
qudits}=&&\cos\theta_2(\cos\theta_1|00\rangle+\sin\theta_1|11\rangle)
+\sin\theta_2(\sin\theta_3\sin\theta_4\cdots
\sin\theta_{d-1}|22\rangle+\sin\theta_3\sin\theta_4\cdots
\cos\theta_{d-1}|33\rangle\nonumber\\
&&+\sin\theta_3\sin\theta_4\cdots
\cos\theta_{d-2}|44\rangle+\cdots+\sin\theta_{3}\cos\theta_{4}|d-2,d-2\rangle+\cos\theta_{3}|d-1,d-1\rangle).
\end{eqnarray}
\end{widetext}
We now choose the unitary transformation matrix of particle A as:
$U(A)=\cos\zeta_a|0\rangle\langle0|+\sin\zeta_ae^{-i\phi_a}|0\rangle\langle1|+\sin\zeta_ae^{i\phi_a}|1\rangle\langle0|
-\cos\zeta_a|1\rangle\langle1|+\sum_{n=2}^{d-1}|n\rangle\langle n|$,
or in the matrix form
\begin{eqnarray}\label{unitary}
U(A)=\left(\begin{matrix}\cos\zeta_a&\sin\zeta_ae^{-i\phi_a}&0&\cdots&0\\
\sin\zeta_ae^{i\phi_a}&-\cos\zeta_a&0&\cdots&0\\
0&0&1&\cdots&0\\
\vdots&\vdots&\vdots&\ddots&\vdots\\
0&0&0&\cdots&1
\end{matrix}\right).
\end{eqnarray}
The matrix $U(B)$ has the same form as $U(A)$. Substitute them into
the CGLMP inequality, and let $\zeta_1=0$, $\zeta_2=\pi/4$,
$\phi_1=0$, $\phi_2=0$, $\varphi_1=0$, $\varphi_2=0$, one obtains
$I_d=\frac{1}{2}(2+\cos2\eta_1-\sin2\theta_1\sin2\eta_1+\cos2\eta_2+\sin2\theta_1\sin2\eta_2)
\cos^2\theta_2+2\sin^2\theta_2\leq(1+\sqrt{1+\sin^2(2\theta_1)})\cos^2\theta_2+2\sin^2\theta_2$.
Similarly, the equal sign occurs at
$\eta_1=-\eta_2=-\tan^{-1}[\sin(2\theta_1)]$. Obviously, since
$1+\sqrt{1+\sin^2(2\theta_1)}$ is larger than 2, as a result, the
maximal value of $I_d$ is larger than $2$. In other words, the CGLMP
inequality is violated for any $\theta_1\neq0$ or $\pi/2$. In the
second and the third step, we have assumed that $\theta_1\neq0$ or
$\pi/2$ (i.e., the coefficients of $|00\rangle$ and $|11\rangle$ are
not zero), which is reasonable because for any entangled two-qudit
state there are at least two nonzero coefficients. Therefore, we can
choose any two of them. For simplicity and convenience, we assume
that the coefficients of $|00\rangle$ and $|11\rangle$ are not zero.
This ends the proof of Gisin's theorem for two qudits.

It is worth mentioning that there are other equivalent simplified
versions of the CGLMP inequality \cite{Gill}, for example,
\begin{eqnarray}\label{gill}
&&P(A_2<B_1)-P(A_2<B_2)-P(B_2<A_1)\nonumber\\
&&-P(A_1<B_1) \le 0,
\end{eqnarray}
where $P(A_2<B_1)$ is understood as $P(A_2=0, B_1=1)+P(A_2=0,
B_1=2)+P(A_2=1, B_1=2)$ when the dimension $d=3$. Following the
similar procedure developed above, one may also complete the proof
of the Gisin's theorem for two qudits based on the elegant
simplified inequality (\ref{gill}).

 Nevertheless, the above Gisin's theorem only indicates that any
pure entangled state of two qudits violates the CGLMP inequality. It
does not give us further information about the maximal quantum
violations of a given state. One notices that the unitary
transformations used in the proof are only $SU(2)$ matrices [see
Eqs. (\ref{qutrit-unitary})(\ref{unitary})], which are only parts of
the full $SU(d)$ transformations. If we apply the full $SU(d)$
transformations to the CGLMP inequality, it is expected that
stronger quantum violations for a given two-qudit state can be
obtained. In this case, generally, it is hard to have an analytical
proof of the Gisin's theorem because of too many parameters involved
in the $SU(d)$ transformations. Instead, we may have a numerical
proof. For instance, in Fig. 1, we have provided a numerical proof
of the Gisin's theorem for two qutrits.

\begin{figure}\label{fig1}
\includegraphics[width=85mm]{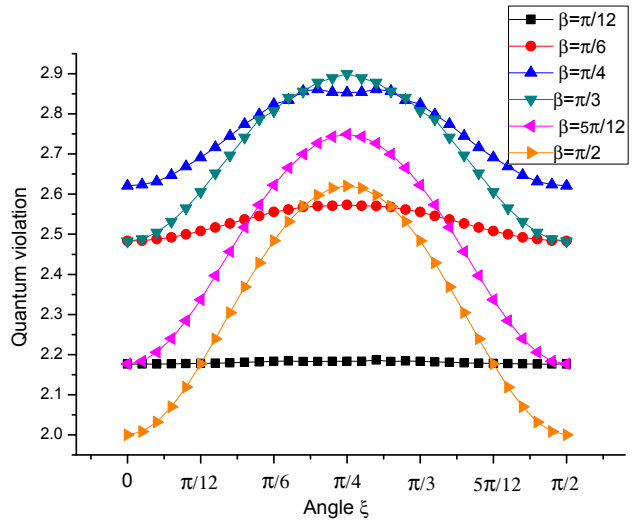}\\
 \caption{(Color online) Numerical proof of the Gisin's theorem for two qutrits. The two-qutrit
 state in the Schmidt-decomposition form reads
 $|\psi\rangle_{\rm qutrits}=\kappa_0 |00\rangle+\kappa_1 |11\rangle+\kappa_2 |22\rangle$,
where $\kappa_0=\sin\beta \cos\xi$, $\kappa_1=\sin\beta \sin\xi$,
$\kappa_2=\cos\beta$. The state $|\psi\rangle_{\rm qutrits}$
 violates the CGLMP inequality for all the parameters $\beta$ and $\xi$ (except the points with $\beta=\pi/2$,
 $\xi=0$ or $\xi=\pi/2$). In the figure we have plotted the curves with $\beta=\pi/12,
 \pi/6,\pi/4,\pi/3,5\pi/12$ and $\pi/2$. One may have an empirical formula numerically fitting the
curves as $I_3^{rough} \simeq 0.5491+0.9344\times {\cal
I}_1^{2.3682}+2.5871\times {\cal I}_2^{-0.031}-2.0636\times{\cal
I}_1^{2.6375}\times {\cal I}_2^{-0.6455}$, where ${\cal
I}_1=\kappa_0^4+\kappa_1^4+\kappa_2^4$ and ${\cal
I}_2=\kappa_0^6+\kappa_1^6+\kappa_2^6$. For any $\beta \ne \pi/2$
and $\xi=0$ (or $\xi=\pi/2$), one has $I_3^{rough}\ge 2$. For
instance, in the case of $\beta=\pi/6$ and $\xi=2\pi/15$, one has
$I_3^{rough} \simeq 2.5366$, which violates the CGLMP inequality.}
\end{figure}

In summary, we have shown analytically that all pure entangled
states of two $d$-dimensional systems violate the CGLMP inequality.
Thus one has the Gisin's theorem for two qudits.
Recently, a coincidence Bell inequality for three three-dimensional
systems (three qutrits) has been proposed (see inequality (4) of
Ref. \cite{ACGKKOZ}). This probabilistic Bell inequality possesses
some remarkable properties: (i) It is a tight inequality; (ii) It
can be reduced to the CGLMP inequality for two-qutrit when the
measurement outcomes of the third observer are set to zero; (iii) It
can be reduced to the Bell inequality for three-qubit based on which
one has the Gisin's theorem for three qubits (see inequality (6) of
Ref. \cite{J.L. Chen-2004}) when each observer's measurement
outcomes are restricted to 0 and 1. Therefore, the Bell inequality
(4) in Ref. \cite{ACGKKOZ} is a very good candidate for proving the
Gisin's theorem of three qutrits. We have randomly chosen thousands
of points for the pure three-qutrit states to find that the Gisin's
theorem for three-qutrit holds. An analytical proof of the Gisin's
theorem for three-qutrit is under development, which we shall
investigate subsequently.


This work is supported in part by NSF of China (Grants No. 10575053
and 10605013), Program for New Century Excellent Talents in
University, and the Project-sponsored by SRF for ROCS, SEM.

\end{document}